# Chiral transport in Compact topological waveguide arrays

Kai Hong[1], Jian Wang[1] and Lin Chen[1,2,*]

1 Wuhan National Laboratory for Optoelectronics and School of Optical and Electronic Information, Huazhong University of Science and Technology, Wuhan 430074, China
2 Shenzhen Huazhong University of Science and Technology Research Institute, Shenzhen 518063, China

* Corresponding author.
Email: chen.lin@mail.hust.edu.cn.

## Abstract

Waveguide arrays have a wide range of applications, including optical communication, topological systems, and quantum computing. Chiral transport has broad prospects in optical computing, optical switching, optical isolation, and polarization control. However, the lengths of chiral transmission devices in waveguide arrays are typically very long due to adiabatic limit. We introduce topological protection in encircling exceptional points systems, enabling the larger gap size between the bulk states and topological edge states (TESs). Thus, the restriction from adiabatic limit on the rapid evolution of TESs is eased, thereby significantly reducing the device size. We experimentally demonstrate that the chiral transport has been achieved in a topological waveguide array that is only 75 μm long. Our research fuses the topology with non-Hermitian physics to develop highly-integrated photonic chips for further advance of nano-photonics.

Waveguide arrays are powerful and highly promising platforms that have attracted widespread attention in the field of optics [1]. In optical waveguide arrays, many interesting functionalities have been achieved, such as Bloch oscillations [2], negative refraction [3], broadband couplers [4], lasers [5], bound states in the continuum [6], non-Hermitian skin effect [7], programmable devices [8], artificial gauge fields [9], synthetic dimensions [10], Floquet engineering [11], and chiral transport [12-17]. Waveguide arrays also offer various other functionalities and applications in topological systems [18] and quantum systems [19]. Chiral transport finds wide applications in quantum computing [15,16,20], asymmetric optical switches [21,22], polarization controllers [23,24], and optical isolators [25], bearing significant importance. Chiral transport is generally implemented in waveguide arrays through methods such as non-Abelian braiding [16], non-Abelian Thouless pumping [13,15], and exceptional points (EPs) encircling [14,17]. However, current devices for chiral transport in waveguide arrays require extremely long lengths to meet adiabatic conditions. The ratios of the device lengths to working wavelength in previous works [14-17] were approximately $6.2\times10^4$, $6.2\times10^4$, $6.2\times10^4$, and $1.5\times10^3$, respectively.

In this Letter, facing the challenge of long device lengths for chiral transport devices in waveguide arrays, we introduce topological protection in EPs encircling systems. There is larger energy gap between topological edge states (TESs) and bulk modes than those in topologically trivial systems, which enables the evolution of TESs to meet adiabatic conditions even in rapid evolution speed. Thus, the length of the chiral transport device in the waveguide arrays is greatly reduced. We experimentally achieved chiral transport of TESs in a 75 μm silicon waveguide array in the communication band around 1550 nm. The ratio of the device length we designed to the working wavelength is 48, at least an order of magnitude lower than previous works[14-17]. Our research demonstrates the robustness and compactness of TES chiral transport in non-Hermitian topological waveguide arrays, which means more interesting physics and applications.

To realize compact chiral TESs transmission, we employed a two-levels non-

Hermitian Rice-Mele model. The unit-cells consist of two sublattices (sublattice 1 and 2) [Fig. 1(a)]. We introduce loss $\gamma$ at sublattice 1 and detuning of onsite energy $\beta$ at sublattice 2 in each unit-cell. The intra-cell and inter-cell coupling coefficients are denoted as $\kappa_1$ and $\kappa_2$ respectively. Firstly, under periodical boundary condition, we try to calculate the topological invariants, i. e. Zak phase $\Phi_{Z,n}$. According to Fourier transformation, such system referring to the non-Hermitian Rice-Mele model can be captured by the following bulk momentum-space Hamiltonian [26]:

$$H(k_x)=\begin{bmatrix} \mathrm{i}\gamma & \kappa_1+\kappa_2\mathrm{e}^{-ik_x} \\ \kappa_1+\kappa_2\mathrm{e}^{\mathrm{ik_x}} & \beta \end{bmatrix} \tag{1}$$

where $k_x$ is the Bloch wave vector. When we considering the condition without detuning and loss, and with coupling coefficient $\kappa_2 = \kappa$, by solving the eigenstates $|\mu_\mathrm{n}\rangle$ of the non-Hermitian bulk momentum-space Hamiltonian, the Zak phase $\Phi_{Z,n}$ can be calculated as $\Phi_{Z,n} = i\int_0^{2\pi}\langle\mu_n(k_x)|\frac{\partial}{\partial k_x}|\mu_n(k_x)\rangle dk_x$ (n=1, 2). We analyze the Zak phase governed by coupling coefficient $\kappa_1$ [Fig. 1(b)]. According to bulk boundary condition [26], the non-zero topological invariant $\Phi_Z = \pi$ predicts the existence of two TESs (TES1 and TES2), when the system is under open boundary condition (OBC). In contrast, $\Phi_Z = 0$ means that the system supports only bulk states. Then, We consider the case with detuning and loss in OBC photonic lattice. This system is composed of N units and Hamiltonian $H_{2\mathrm{N}\times2\mathrm{N}}$ can be represented as following:

$$H_{2\mathrm{N}\times2\mathrm{N}} = \begin{bmatrix} i\gamma & \kappa_1 & & & & & \\ \kappa_1 & \beta & \kappa_2 & & & & \\ & \kappa_2 & i\gamma & \cdots & & & \\ & & & \cdots & & & \\ & & & \cdots & \beta & \kappa_2 & \\ & & & & \kappa_2 & i\gamma & \kappa_1 \\ & & & & & \kappa_1 & \beta \end{bmatrix}. \tag{2}$$

The real part of the eigenvalues of the Hamiltonian $H_{2\mathrm{N}\times2\mathrm{N}}$ corresponds to the propagation constants of the respective eigenstates. The propagation constants of all states, varying with detuning $\beta$ and loss $\gamma$, form a Riemann surface, as illustrated in Fig. 1(c) and Fig. 1(d). When $\kappa_1 > \kappa_2$ $(\Phi_Z = 0)$, all states of the system are bulk states, and there is N EPs [Fig. 1(c)]. When $\kappa_1 < \kappa_2$ $(\Phi_Z = \pi)$, this system exhibits 2 TESs and N EPs, with the TESs degenerating into the first EP, and 2N-2 bulk modes degenerating into the last N-1 EPs [Fig. 1(d)]. When loss and detuning are introduced, the Zak phase is no longer quantized and cannot describe the topological phase of the system accurately. We can only roughly determine the topological phase of the system based on the characteristics of the energy spectrum of TESs and bulk states. We observe that when the loss $\gamma$ is significantly large, the real parts of the eigenvalues of TESs and bulk modes degenerate, which means that topological protection is invalid [27]. In general, the tiny detuning $\beta$ only causes a shift in the eigenvalues of the system without causing topological phase transition (TPT) [28]. As a result, TPT has occurred when sufficient loss causes the overlap of TESs with the bulk states in the energy spectrum. When the evolution trajectories of system parameters encircle only the first EP in the topological phase on the Riemann surface, the system not only exhibits robustness but also more easily satisfies the adiabatic condition, allowing more compact device design. This is due to the adiabatic condition

$$\mathrm{U}_{\mathrm{m,m'}} = \frac{\langle \psi_{\mathrm{m'}} | \frac{\partial}{\partial \mathrm{z}} | \psi_{\mathrm{m}} \rangle}{|\mathrm{Re}(E_{\mathrm{m'}} \text{-} E_{\mathrm{m}})|} \ll 1 \tag{3}$$

where $\mathrm{U}_{\mathrm{m,m'}}$ is the adiabatic factor, the Dirac notation $\langle \psi_{\mathrm{m'}} |$ and $| \psi_{\mathrm{m}} \rangle$ refer to two different eigenstates of our system. $E_{\mathrm{m}}$ and $E_{\mathrm{m'}}$ are the corresponding eigenvalues (m' $\neq$ m), $z$ represents the propagation direction. When the system only supports bulk states, the energy gaps $|\mathrm{Re}(E_{\mathrm{bulk'}} \text{-} E_{\mathrm{bulk}})|$ between them are very small [Fig. 1(c)]. In this scenario, to satisfy adiabatic conditions, the evolution speed of the transmission state $\partial | \psi_{\mathrm{bulk}} \rangle / \partial z$ need to be very slow, corresponding to longer device structures. When the

system supports TESs, the energy gaps $|\mathrm{Re}(E_{\mathrm{TESs}}\text{-}E_{\mathrm{bulk}})|$ between the TESs and bulk states will become smaller as the position ($\beta/\kappa$, $\gamma/\kappa$) is further away from the origin in parameter space $\beta$O$\gamma$, corresponding to longer device structures [Fig. 1(d)]. When we devise dynamical processes to achieve chiral TESs transmission, the trajectories encircling EPs formed by bulk modes in the parameter space will inevitably lead the system falling into the trivial phase [Fig. 1(e)]. Therefore, for the chiral TESs transmission device to be robust and more compact, the evolution trajectories in parameter space need to be carefully designed to remain in the non-trivial topological phase. It is crucial to encircle the first EP formed by TESs to ensure the system remaining in the topological phase, for example, the red evolution trajectory in Fig. 1(e).

To demonstrate compact chiral TESs transmission in the Riemann surfaces, we constructed a system comprised of topological waveguide arrays [Fig. 2(a)]. This system is composed of 3 unit cells and each unit cell consists of two waveguides (waveguide 1, waveguide 2). Such a system is fabricated on a silicon-on-isolator (SOI) wafer with a top silicon layer of 220 nm and a $SiO_2$ covered layer of 1 μm. A 20-nm-thick chromium layer with varied widths is placed on the waveguide 1 in each unit cell to introduce a position-dependent loss. According to the coupled-mode theory, the topological waveguide arrays system state $|\psi\rangle$ propagating along $z$ follows a Schrödinger-type equation $i\partial/\partial z|\psi\rangle = H_{6\times6}|\psi\rangle$, where $H_{6\times6}$ is the Hamiltonian for the E.Q. (2) with 2N=6. The six eigenvalues of $H_{6\times6}$ are $E_{1-6} = (\beta + i\gamma)/2 \pm \sqrt{(\beta - i\gamma)^2/4 + R_{1-3}}$ [Re($E_{\mathrm{m}}$)<Re($E_{\mathrm{n}}$), m<n, Re($E$) denotes the real part of the eigenvalue $E$], here $R_{1-3}$ are the three positive real roots of the equation : $x(x-\kappa_1^2-\kappa_2^2)^2 - \kappa_1^2(x-\kappa_1^2)^2 = 0$ and according to Rolle's theorem, $0 < R_1 < \kappa_1^2 < R_2 < \kappa_1^2 + \kappa_2^2 < R_3$ (see Supplemental Material, Note 1 for the details of the eigenvalues). TES1(TES2) correspond to eigenvalue $E_3$ ($E_4$) and eigenstate $|\psi_3\rangle(|\psi_4\rangle)$. When $\beta = 0$ and $\gamma = 2\sqrt{R_1}$ , eigenvalues ($E_3$, $E_4$) and eigenstates ($|\psi_3\rangle$, $|\psi_4\rangle$) are degeneracy as an EP. We show the calculated eigenvalues of the Hamiltonian in the $(\beta/\kappa_2, \gamma/\kappa_2)$ parameter space in Fig. 2(b)-2(e). This evolution trajectories are meticulously designed to encircle an EP to realize chiral TESs transmission. We map the evolution trajectory in Fig. 2(b)-2(e) to the topological waveguide array structure. The essential elements in determining the mode evolution are $\beta$, $\gamma$, $\kappa_1$ and $\kappa_2$, which are controlled by the waveguide width difference $\Delta d = d_1 - d_2$, chromium width $d_{Cr}$ and gap distance $g_1$, $g_2$. In section A$\rightarrow$D, $\kappa_2/\kappa_1$ remains unchanged and $\beta$ changes like a periodic sine function. In section B$\rightarrow$C, $\gamma$ increases first and then decreases; $\kappa_1$ and $\kappa_2$ decrease first and then increase. The waveguide widths ($d_1$, $d_2$), the gap distances ($g_1$, $g_2$), the chromium width ($d_{Cr}$) and Hamiltonian parameters, varying along the *z* direction, are shown in Fig. S1 (see Supplemental Material, Note 2). For the designed device, the adiabaticity factor $U \ll 1$ throughout the entire transmission process when the length is 75 μm (see Supplemental Material, Note 3). Due to the large energy gap between the TESs and the bulk states, there is almost no excitation of bulk states during the transmission process of the TESs even for a short device length of 75 μm. When TES1 (TES2) passes through the intersection of the self-intersecting Riemann surface, it transforms into TES2 (TES1), corresponding to the $U = +\infty$ in Fig. 1(f) [Fig. 1(g)]. For convenience, we only consider TESs ($|\psi_3\rangle$, $|\psi_4\rangle$) to describe the transmission process of TESs and we define $|\psi_{34}\rangle$ ($|\psi_{43}\rangle$) is $|\psi_3\rangle$($|\psi_4\rangle$) first and then $|\psi_4\rangle$($|\psi_3\rangle$) when the evolution trajectory through intersection of Riemann surface. Input TESs from the left port corresponds to dynamic Hamiltonian trajectory anti-clockwise (ACW) encircling an EP [Figs. 2(b), 2(c)]. TES2 and TES1 input corresponds to Figs. 2(b), 2(c), respectively. For TES2 input, the initial state $|\psi_4\rangle$ at the starting point $(\beta/\kappa_2, \gamma/\kappa_2) = (\beta_0, 0)$ is located on the upper sheet of the Riemann surface [Fig. 2(b)]. When the absolute value of $\beta_0$ is close to zero, the field of TESs cannot be located in an edge waveguide but forms symmetric and antisymmetric states. Therefore, we set $\beta_0$ to $0.48\kappa_2$ . The state $|\psi_{43}\rangle$ is always dominant and low loss as the imaginary part of the eigenvalue of $|\psi_{43}\rangle$ is always close to zero. Though a small $|\psi_{34}\rangle$ is excited since adiabaticity can not be strictly fulfilled, but its contribution is small and it is further attenuated as the imaginary part of the eigenvalue of $|\psi_{34}\rangle$ is much larger. The Hamiltonian finally returns to ($\beta_0$ , 0) and the output state is dominated by $|\psi_3\rangle$ on the lower sheet of the Riemann surface. For TES1 input, the initial state $|\psi_3\rangle$ at the starting point $(\beta/\kappa_2, \gamma/\kappa_2)$ = ($\beta_0$ , 0) is located on the lower sheet of the Riemann surface [Fig. 2(c)]. The state $|\psi_{34}\rangle$ evolves slowly to B, where the dominant eigenstate is $|\psi_{34}\rangle$ and few eigenstate $|\psi_{43}\rangle$ . In section BC, the dominant eigenstate $|\psi_{34}\rangle$ suffers from high loss and it is completely dissipated. In contrast, $|\psi_{43}\rangle$ is low loss and becomes dominant at C, i.e., a NAT occurs. The final state returns to $|\psi_3\rangle$ at ($\beta_0$ , 0) on the lower sheet of the Riemann surface. Input TESs from the right port corresponding to dynamic Hamiltonian trajectory clockwise (CW) encircling an EP shows in Figs. 2(d), 2(e). In the same way, The final state returns to $|\psi_4\rangle$ at ($\beta_0$ , 0) on the upper sheet of the Riemann surface regardless of $|\psi_3\rangle$ or $|\psi_4\rangle$ input. It should be noted that the output states for CW and ACW loops are always TES2 and TES1, respectively, regardless of the input TESs.

In order to clearly show chiral TESs transmission, we performed full-wave simulations using finite-difference time-domain methods (Fig. 3). The field intensity of TES1 and TES2 is mainly located in the waveguide1 and waveguide2 at the edge of the waveguide arrays, respectively. The TES1 and TES2 input from the left port (Fig. 3 top-left), or the right port (Fig. 3 bottom-right). The TES2 and TES1 output from the right and left ports, respectively, regardless of whether TES1 or TES2 is input. The TESs purity in the final state is a key quantity. The TESs purity $\eta$ is TES-to-total power ratio in the final state and $\eta_m = E_{TESm}/E_{total}$ (m=1, 2), where $E_{TESm}$ and $E_{total}$ correspond to the TESm energy and the final state energy in output port, respectively. $\eta$ and $\eta'$ correspond to left and right input, respectively. When TES1 (TES2) input, $\eta_1, \eta_2, \eta_1'$ and $\eta_2'$ are 99% (97%), 1% (3%), 1% (2%), 90% (98%), respectively (see Supplemental Material, Note 4 for the details of the evolution process). Unlike the strategy that the evolution trajectories always stay in the topological phase, we also conducted simulation experiments for the case of encircling multiple EPs. We set the same device length L=75 μm, and the control parameter evolution trajectory encircles multiple EPs on the Riemann surface. We found that the TESs purity $\eta_1, \eta_2, \eta_1'$ and $\eta_2'$ are 4% (78%), 0.2% (6%), 10% (0.1%), 84% (73%), respectively, when TES1 (TES2) input. The device encircling multiple EPs cannot achieve TESs chiral transmission, as there are numerous bulk states in the output modes, preventing TES from dominating, unless the device length is significantly increased to reduce the interference from bulk states. Previous works on EPs in waveguide arrays have employed very long device structures. For instance, in Ref. [14], the device length is 50 mm at a wavelength of 808 nm, and in Ref. [17], the device length is 2500 μm at a wavelength of 1550 nm. The device we design are robust against the fabrication imperfections. We introduce the waveguide width deviation Δd and the gap distance deviation Δg to the proposed structures. We performed a simulation to verify robustness with Δd and Δg varying from −40 to 40 nm. The device can still achieve chiral TESs transmission (see Supplemental Material, Note 5 for the details of the robustness against the fabrication imperfections).

A measured scanning electron microscope (SEM) image of the fabricated topological waveguide arrays sample is shown in Figs. 4(a)-4(e), where the zoomed-in images in Fig. 4(b)-4(e) represent the region bounded by the same color rectangle in Fig. 4(a), respectively (the fabrication details can be found in Supplemental Material, Note 6). The section of topological waveguide array is shown in Fig. 4(b), Fig. 4(c) and the section of adiabatic waveguide is shown in Fig. 4(d), Fig. 4(e) (see Supplemental Material, Note 7 for the details of Adiabatic waveguides). Grating couplers were placed on both sides of topological waveguide arrays for transmission measurement (see Supplemental Material, Note 8 for the details of the measurement scheme). Both simulated and experimental results of chiral TESs transmission in Fig. 4(f), 4(g) for the TES1 input, and in Fig. 4(h), 4(i) for the TES2 input, respectively. The measurable bandwidth is limited by the operation wavelength range of the laser. $T_{\text{mn}}$ ( $T_{\text{mn}}'$ ) represents the transmission efficiency of the TESm output at the right (left) port of the device when TESn inputs from the port at the left (right). For TES1 or TES2 input from left port (ACW), based on the simulation and experimental results presented in Figs. 4(f)-4(i), where $T_{11} > T_{21}$ and $T_{12} > T_{22}$, we can conclude that the TES1 dominates in the final state. On the contrary, for TES1 or TES2 input from right port (CW), the TES2 dominates in the final state. It clearly indicates chiral TESs transmission. The experimental results show some deviation from simulations, which can be attributed to the fabrication error arising from etching roughness precision and the residual smudges during the fabrication process. The device length can be further reduced, but it may sacrifice the output TES purity and the bandwidth. We considered both the bandwidth and the output TES purity, and set the device length to 75 μm.

In conclusion, we have realized an asymmetry transmission of TESs around EPs in a non-Hermitian Rice-Mele model. Compared to previous works on encircling EPs, our work introduced topological protection on states. Thus, our systems exhibit high robustness to structural parameters, which has the potential to solve the fabrication challenge in photonic integration. At the same time, we find that topology also brings compactness to the chiral TESs transmission. We also show the dynamics of TESs with the time-varying non-Hermitian Hamiltonian, which is quite from the works focusing on the static topological features in non-Hermitian systems. The principles of our TESs asymmetry transmission can be extended to enrich the novel methods to develop on-chip nanophotonic device, such as mode switcher, routers and multiplexer

/demultiplexer. Our research work combines the topological photonics with EP dynamics and may benefit to the basic study on non-Hermitian topology at the platform of thermology, acoustics, electronics, and condensed matter physics.

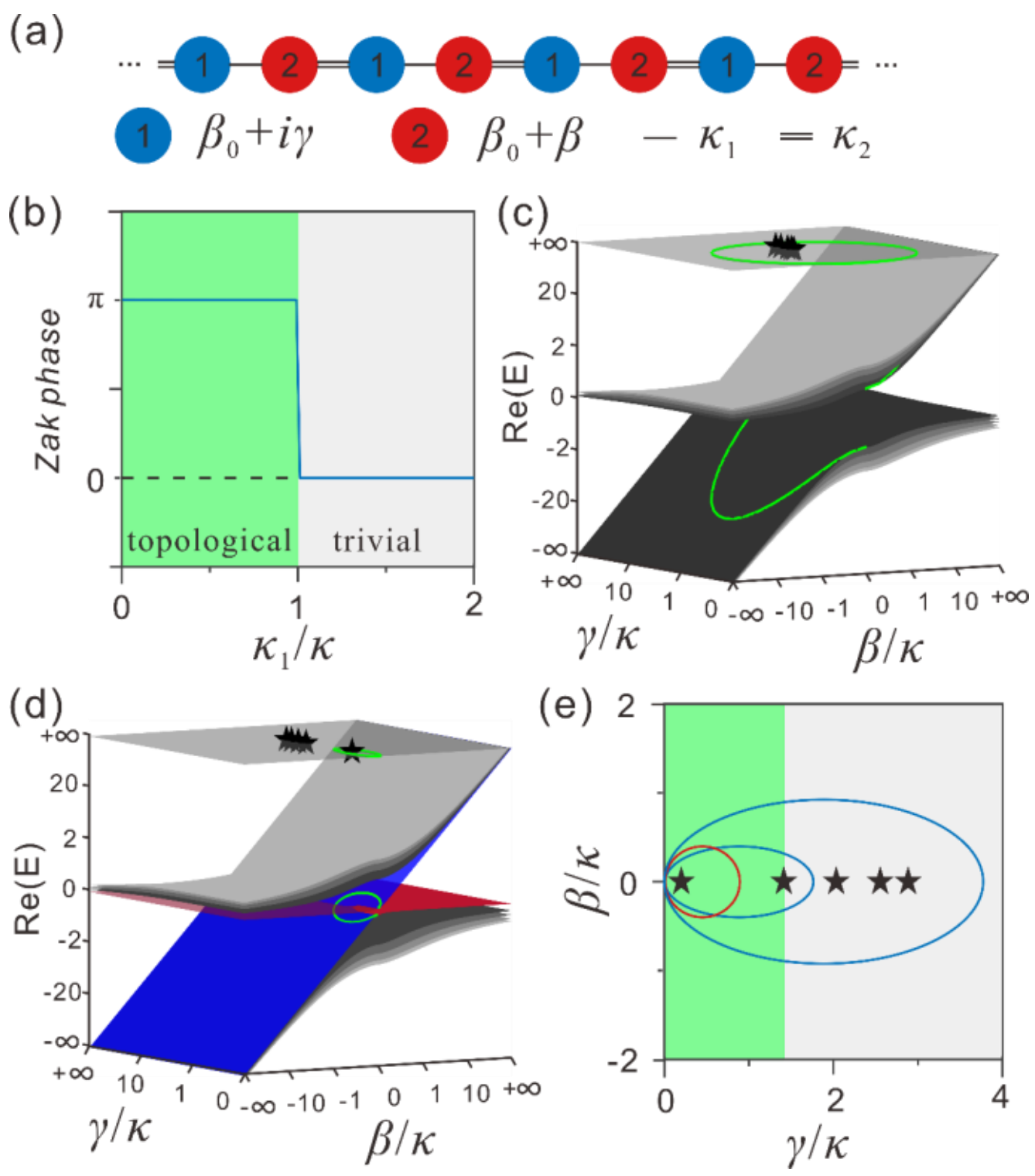

**Fig. 1 The non-Hermitian Rice-Mele system.** (a) The non-Hermitian Rice-Mele model. (b) The Zak phase varies with coupling coefficient $\kappa_1$, and $\kappa_2=\kappa$, $\gamma=0$, $\beta=0$. (c), (d) The real part of the eigenvalues varies with the $\beta$ and $\gamma$ when 2N=10, $\kappa_1=\kappa$, $\kappa_2=0.5\kappa$ (c) and $\kappa_1=0.5\kappa$, $\kappa_2=\kappa$ (d). The colored surfaces represent the eigenvalues corresponding to the TESs, while the gray surfaces represent the eigenvalues corresponding to bulk states. The EPs and the evolution trajectory are projected onto the upper plane. (e) The trajectory of parameters $\beta$ and $\gamma$ encircling an EP. The red curve represents the evolution trajectory entirely in the topological phase, while the blue curves indicate the trajectories that fall into the trivial phase during the evolution process. The green background represents the topological phase while the gray background refers to the trivial phase in (b) and (e). Pentagon markers indicate EPs.

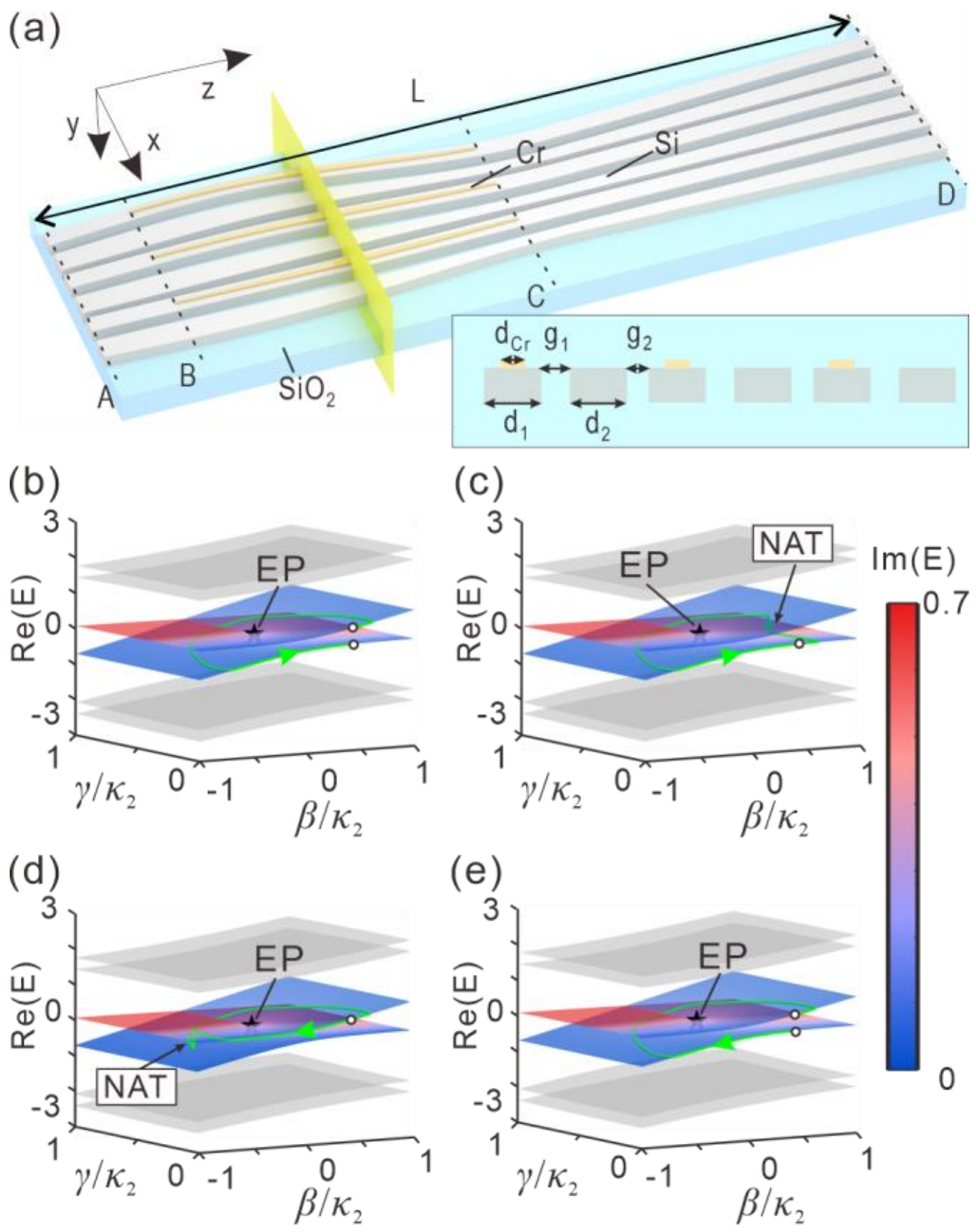

**Fig. 2 The dynamically evolving system of TESs.** (a) Topological silicon waveguide arrays on SOI wafer. The vignette in the lower right corner corresponds to the yellow cross section in the structure. L = 75 μm. (b), (c) ACW and (d), (e) CW loops around an EP in the Riemann surfaces formed by the energy spectra of $H$. Initial state is TES2 in (b), (d), and TES1 in (c), (e). Color surface consists of the eigenvalues of the TESs, and gray surface consists of the eigenvalues of the bulk states. The empty circles present the start and end of the loops. The pentagrams indicate EPs.

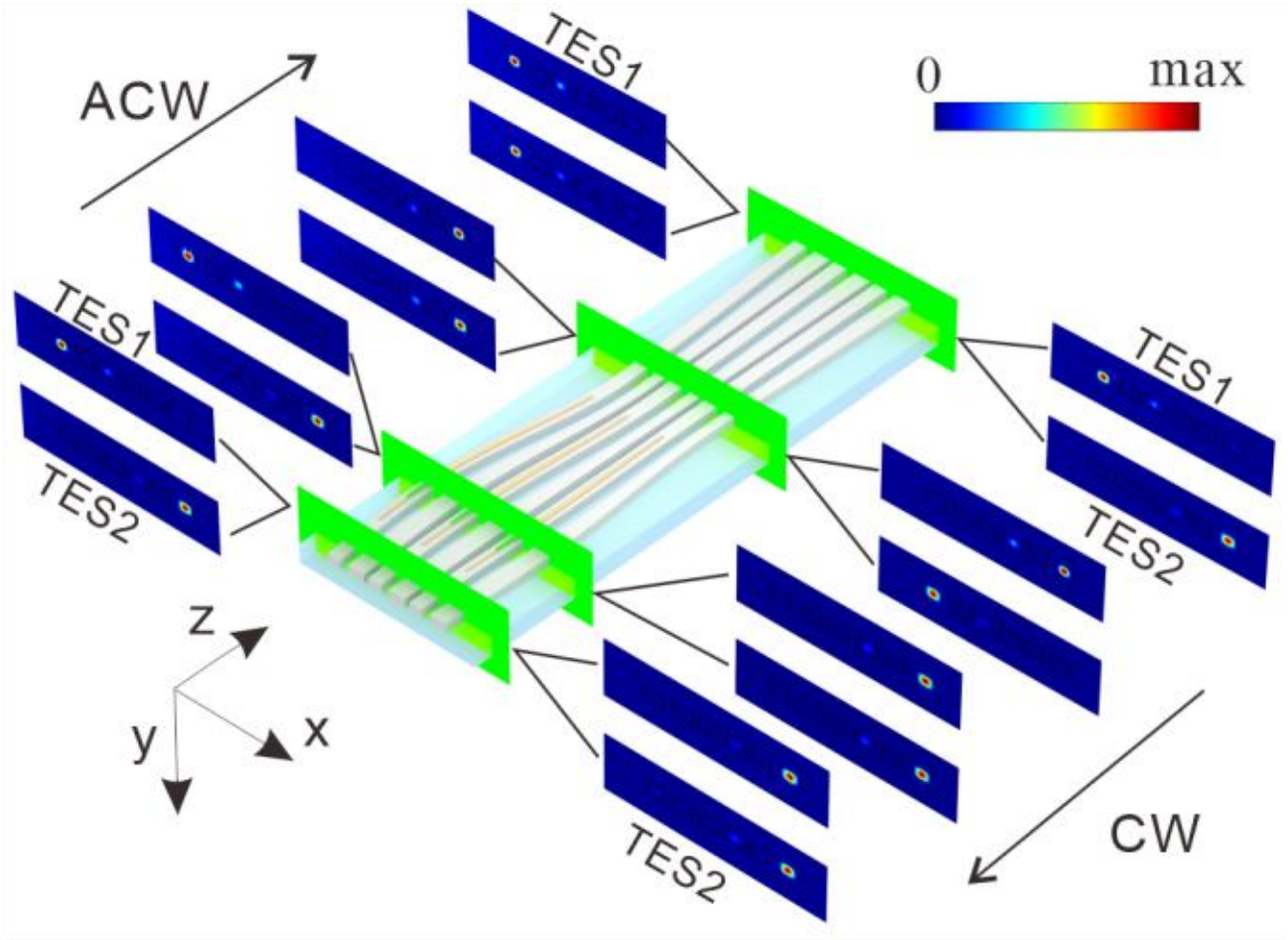

**Fig. 3. Simulated transmission.** The cross-sectional field intensity distributions of $|E|^2$ at 1550 nm marked by several green planes are displayed. ACW and CW correspond to top-left and bottom-right, respectively. TES1 input scenario is plotted above, and TES2 input scenario is plotted below. The waveguide outlines are delineated by black curves.

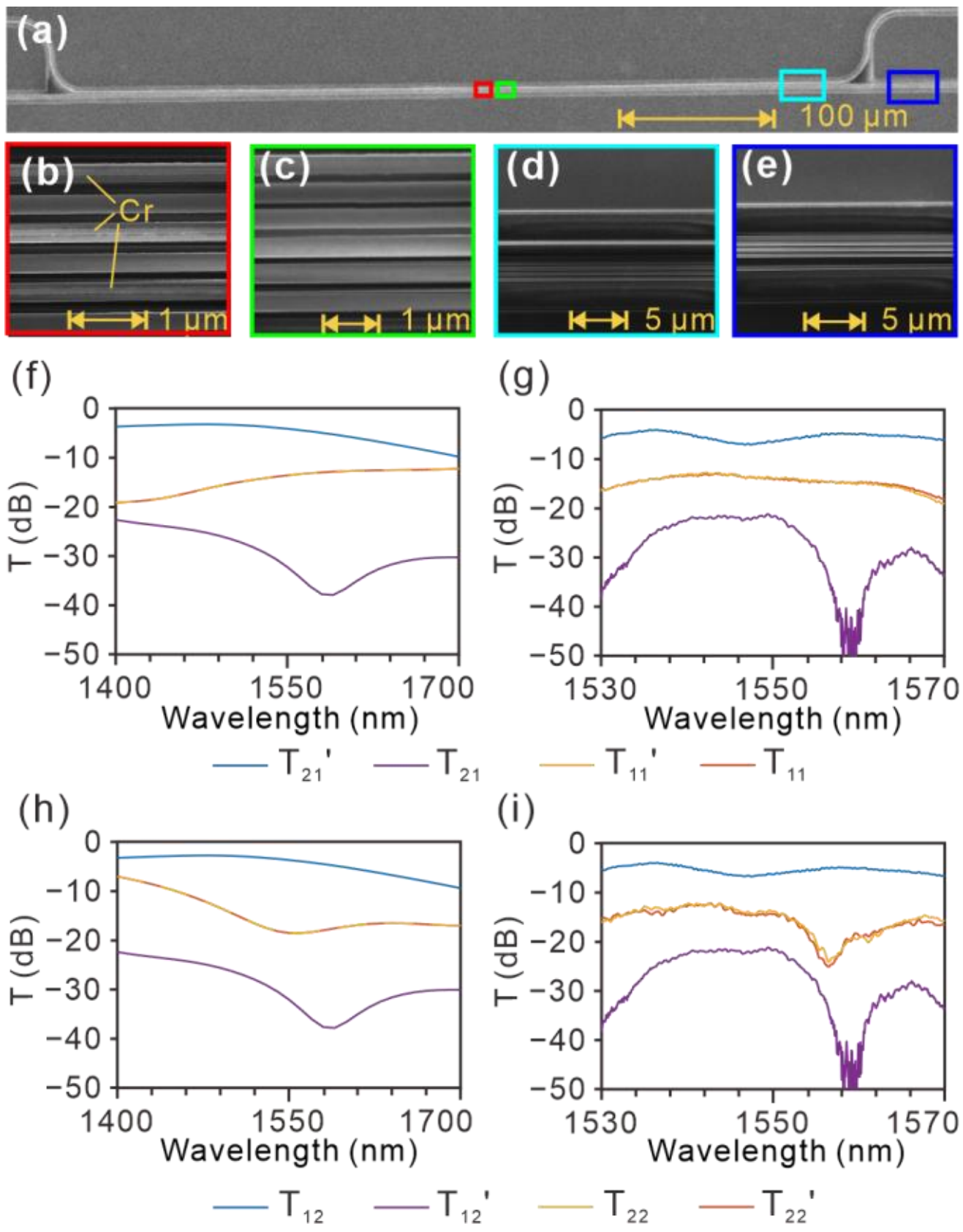

**Fig. 4 Experimental demonstration.** (a) SEM image of the device. (b)-(e) Zoom-in SEM images bounded by the same color rectangles are marked in (a). (f)- (i) Simulated (f), (h) and experimental (g), (i) transmittance spectra for the output ports when TES1 (f), (g) and TES2 (h), (i) input over the wavelength range of 1530–1570 nm and 1400–1700 nm, respectively.

Supplementary material for

# Compact chiral topological edge states transmission by encircling a selected exceptional point

Kai Hong[1], Jian Wang[1] and Lin Chen[1,2,]*

1 Wuhan National Laboratory for Optoelectronics and School of Optical and Electronic Information, Huazhong University of Science and Technology, Wuhan 430074, China
2 Shenzhen Huazhong University of Science and Technology Research Institute, Shenzhen 518063, China

* Corresponding author.

Email: chen.lin@mail.hust.edu.cn.

**Supplementary Note 1: The eigenvalues of sixth-order Hamiltonian**

Our system can be described by the Hamiltonian:

$$H(z)=\begin{bmatrix} i\gamma(z) & \kappa_1(z) & & & & \\ \kappa_1(z) & \beta(z) & \kappa_2(z) & & & \\ & \kappa_2(z) & i\gamma(z) & \kappa_1(z) & & \\ & & \kappa_1(z) & \beta(z) & \kappa_2(z) & \\ & & & \kappa_2(z) & i\gamma(z) & \kappa_1(z) \\ & & & & \kappa_1(z) & \beta(z) \end{bmatrix} \tag{S1}$$

The eigenvalues $E$ satisfy the following equation:

$$|H-E|=0 \tag{S2}$$

The equation S2 can be written as

$$(E-i\gamma)(E-\beta)[(E-i\gamma)(E-\beta)-\kappa_1^2-\kappa_2^2]^2-\kappa_1^2[(E-i\gamma)(E-\beta)-\kappa_1^2]^2=0 \tag{S3}$$

Defining $x=(E-i\gamma)(E-\beta)$, Eq. S3 can be rewritten as

$$x(x-\kappa_1^2-\kappa_2^2)^2-\kappa_1^2(x-\kappa_1^2)^2=0 \tag{S4}$$

The solutions of Eq. S4 are

$$\begin{aligned} R_1 &= \frac{3\kappa_1^2+2\kappa_2^2}{3}-\frac{\sqrt[3]{2}(-6\kappa_1^2\kappa_2^2-\kappa_2^4)}{3X}+\frac{X}{3\sqrt[3]{2}} \\ R_2 &= \frac{3\kappa_1^2+2\kappa_2^2}{3}+\frac{(1+i\sqrt{3})(-6\kappa_1^2\kappa_2^2-\kappa_2^4)}{3\sqrt[3]{4}X}-\frac{(1-i\sqrt{3})X}{6\sqrt[3]{2}} \\ R_3 &= \frac{3\kappa_1^2+2\kappa_2^2}{3}+\frac{(1-i\sqrt{3})(-6\kappa_1^2\kappa_2^2-\kappa_2^4)}{3\sqrt[3]{4}X}-\frac{(1+i\sqrt{3})X}{6\sqrt[3]{2}} \end{aligned} \tag{S5}$$

where $X=\sqrt[3]{9\kappa_1^2\kappa_2^4-2\kappa_2^6+3\sqrt{3}\sqrt{-32\kappa_1^6\kappa_2^6-13\kappa_1^4\kappa_2^8-4\kappa_1^2\kappa_2^{10}}}$ , Eq. S3 can be rewritten as

$$[(E-i\gamma)(E-\beta)-R_1][(E-i\gamma)(E-\beta)-R_2][((E-i\gamma)(E-\beta)-R_3]=0 \tag{S6}$$

Therefore, $E_{1-6}=(\beta+i\gamma)/2\pm\sqrt{(\beta-i\gamma)^2/4+R_{1-3}}$

## Supplementary Note 2: Structural and Hamiltonian parameters

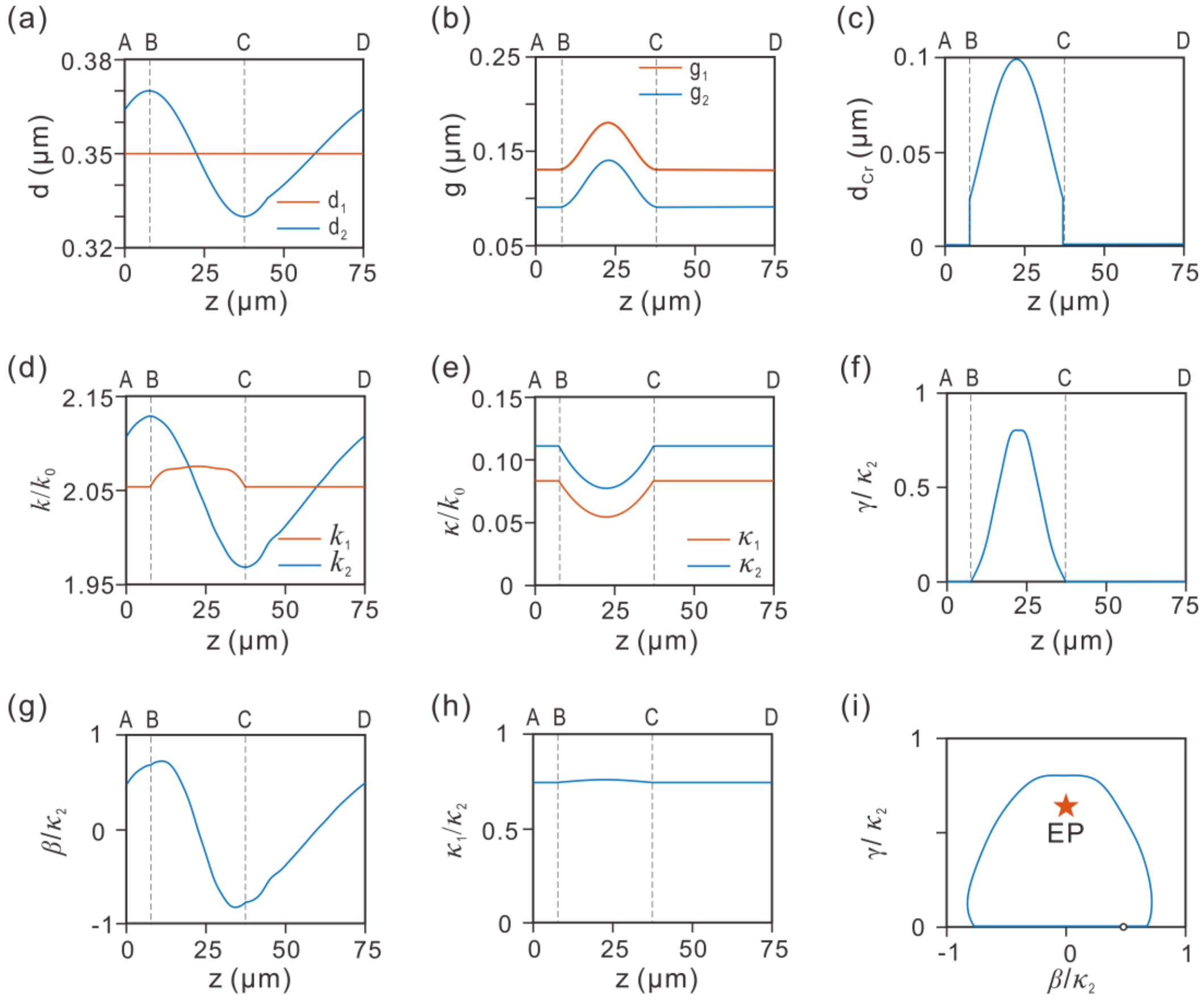

**Fig. S1.** Structural and Hamiltonian parameters. (a-c) The waveguide widths ($d_1$, $d_2$), the gap distances ($g_1$, $g_2$) and the chromium width ($d_{Cr}$), varying along the z direction, respectively. (d-h) The propagation constants $k$ (d), the coupling coefficient $\kappa$ (e), loss rate $\gamma$ (f), the detuning $\beta$ (g), and the ratio of the coupling coefficient $\kappa_1 / \kappa_2$ (h) on the spatial coordinate z at 1550 nm. $k_0$ is the vacuum wave vector and $\beta = k_2 - k_1$. (f) The evolution trajectory. The pentagram indicates an EP point and the empty circles present the start and end of the loops.

In section A→D, the waveguide width $d_2$ varies sinusoidally, and the gap distances changes synchronously. $\kappa_2 / \kappa_1$ remains unchanged and $\beta$ changes varies sinusoidally. In section B→C, the gap distance $g_1$, $g_2$ and the chromium width $d_{Cr}$ increase first and then decrease. Therefore, $\gamma$ increases first and then decreases; $\kappa_1$ and $\kappa_2$ decrease first and then increase. $k_1$ becomes larger due to the effect of the chromium on waveguide1 [Fig. S1(a)-S1(h)]. The evolution process ensures the occurrence of NAT in B-C interval. The evolution trajectory is in the $(\beta/\kappa_2, \gamma/\kappa_2)$ parameter surface [Fig. S1(i)].

## Supplementary Note 3: Adiabatic factor

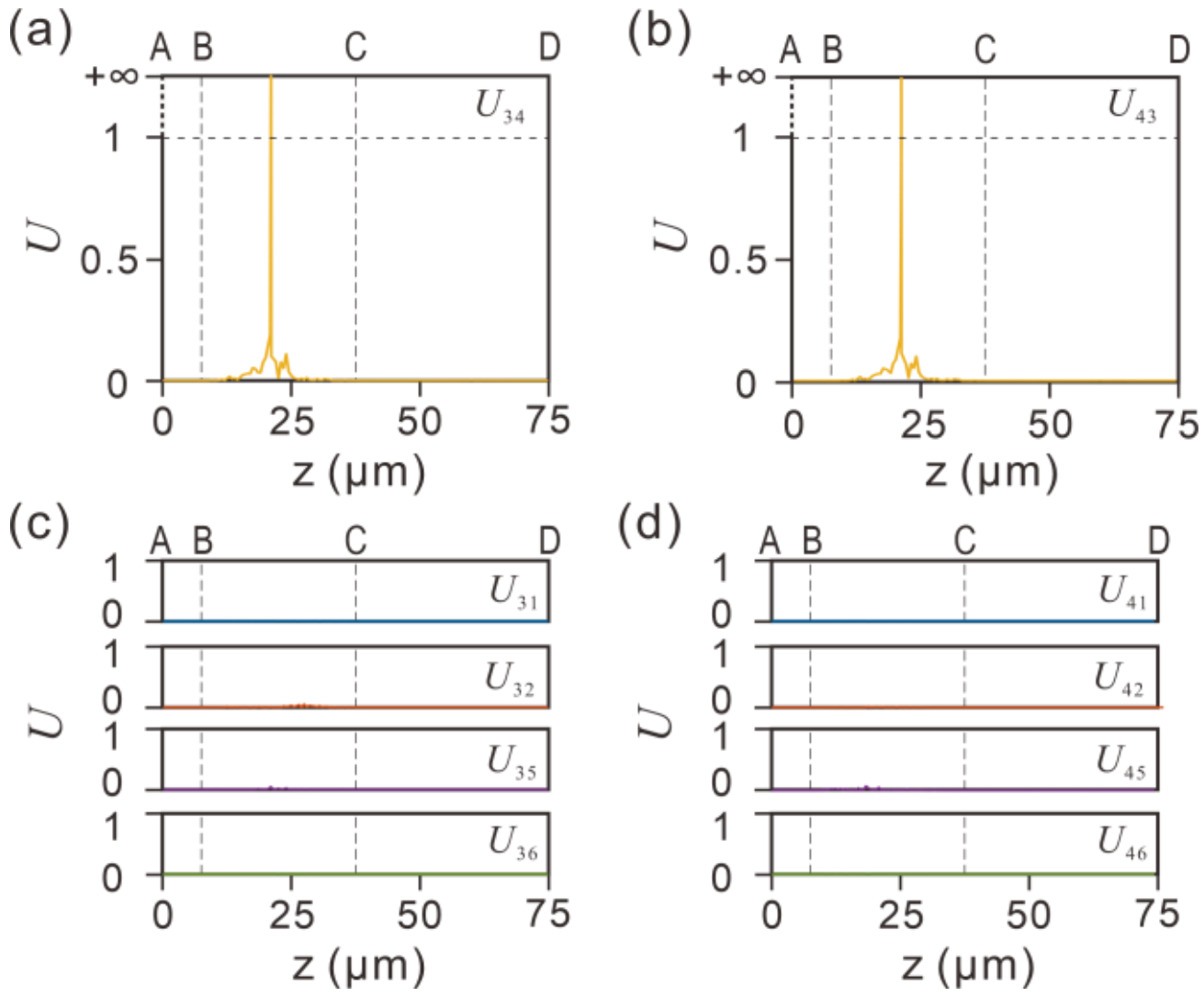

**Fig. S2.** Adiabatic factor. (a)-(d) Adiabatic factor $U$ corresponding to the device structure during TES1 (a), (c) and TES2 (b), (d) transmission.

## Supplementary Note 4: The simulation of evolution process

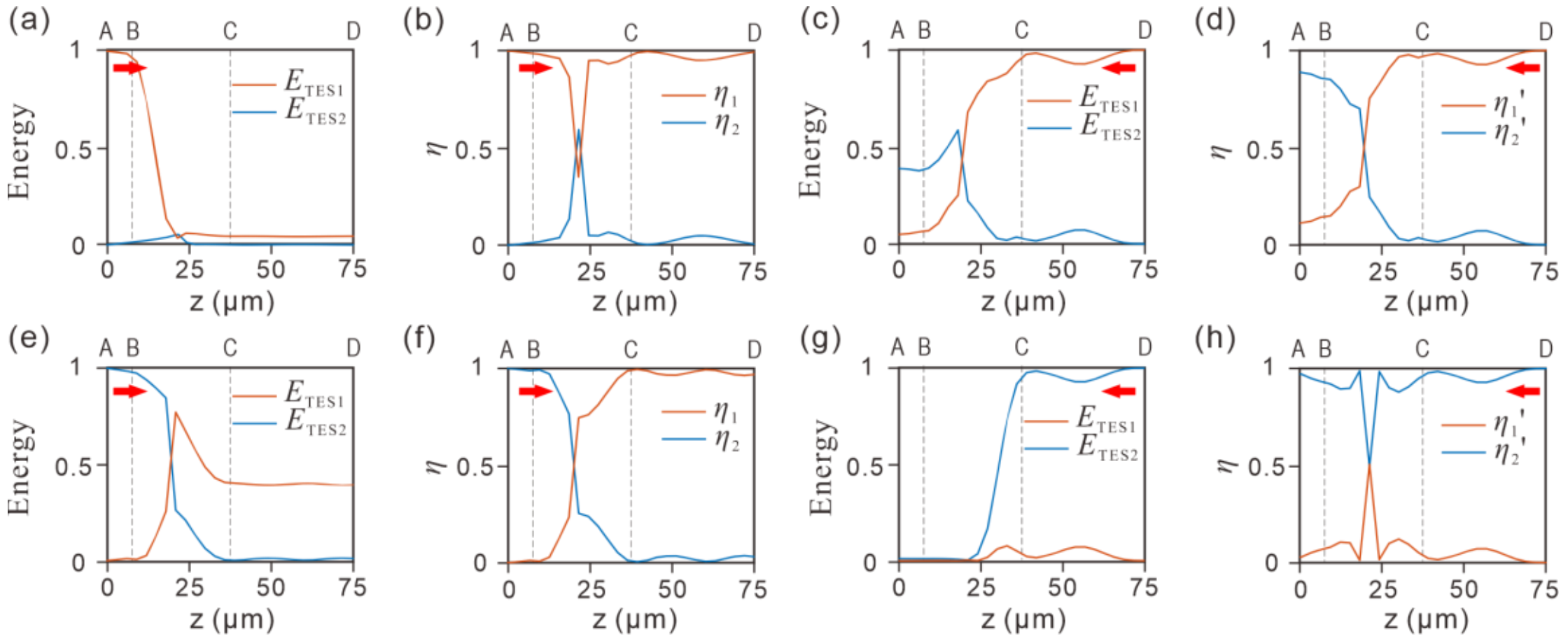

**Fig. S3.** The simulation of evolution process. (a-d) The normalized energy of TESs, $E_{\mathrm{TES1}}$, $E_{\mathrm{TES2}}$, and TESs purity, $\eta$ versus the propagation distance, $z$, when the initial state is TES1 and inputs from the left port (a, b) and the right port (c, d). (e-h) The normalized energy of TESs, $E_{\mathrm{TES1}}$, $E_{\mathrm{TES2}}$, and TESs purity, $\eta$ versus the propagation distance, $z$, when the initial state is TES2 and inputs from the left port (e, f) and the right port (g, h). The normalized energy of TESs defined as $E_{\mathrm{TESm}} = E_m / E_0$, where $E_m$ is the energy of TESm and $E_0$ is the total energy of the initial input.

In order to further demonstrate the asymmetric TESs transmission as indicated by Fig. 2 in the main text, we carried out simulation experiments. We monitor the normalized energy of TESs and TESs purity every 3 μm along the propagation direction z (Fig. S3). For the TES1 input from the left port, the occurrence of NAT results that TES2 dominates. Then, TES1 and TES2 will be reversed when the evolution path passes through the intersection of Riemann surface. Therefore, TES1 dominates in the final state [Figs S3(a)-S3(b)]. For the TES1 input from the left port, however, NAT will not occur and TES2 dominates in the final state [Figs S3(c)-S3(d)]. For the same reason, TES1 (TES2) dominates in the final state when the TES1 inputs from the left (right) port [Figs S3(e)-S3(h)]. The final state is always dominted by TES1 for the TESs input from the left port and TES2 for the TESs input from the right port.

## Supplementary Note 5: Robustness against the fabrication imperfections

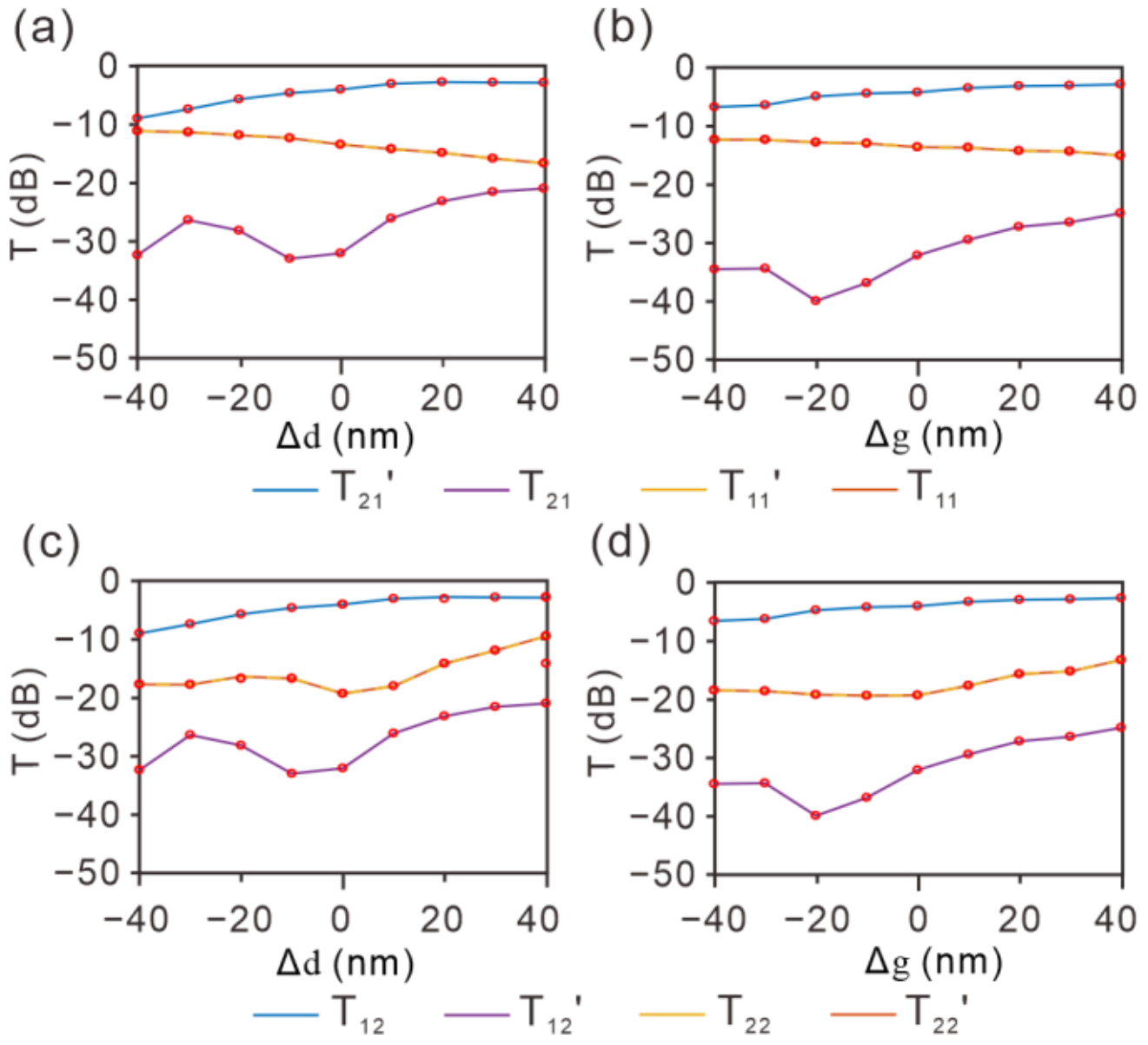

**Fig. S4.** Robustness against the fabrication imperfections. (a), (b) Simulated transmission for TES1 input when (a) the waveguide width deviation Δd and (b) the gap distance deviation Δg vary from −40 to 40 nm. (c), (d) Simulated transmission for TES2 input when (c) the waveguide width deviation Δd and (d) the gap distance deviation Δg vary from −40 to 40 nm.The wavelength is fixed at 1550 nm.

## Supplementary Note 6: Fabrication details

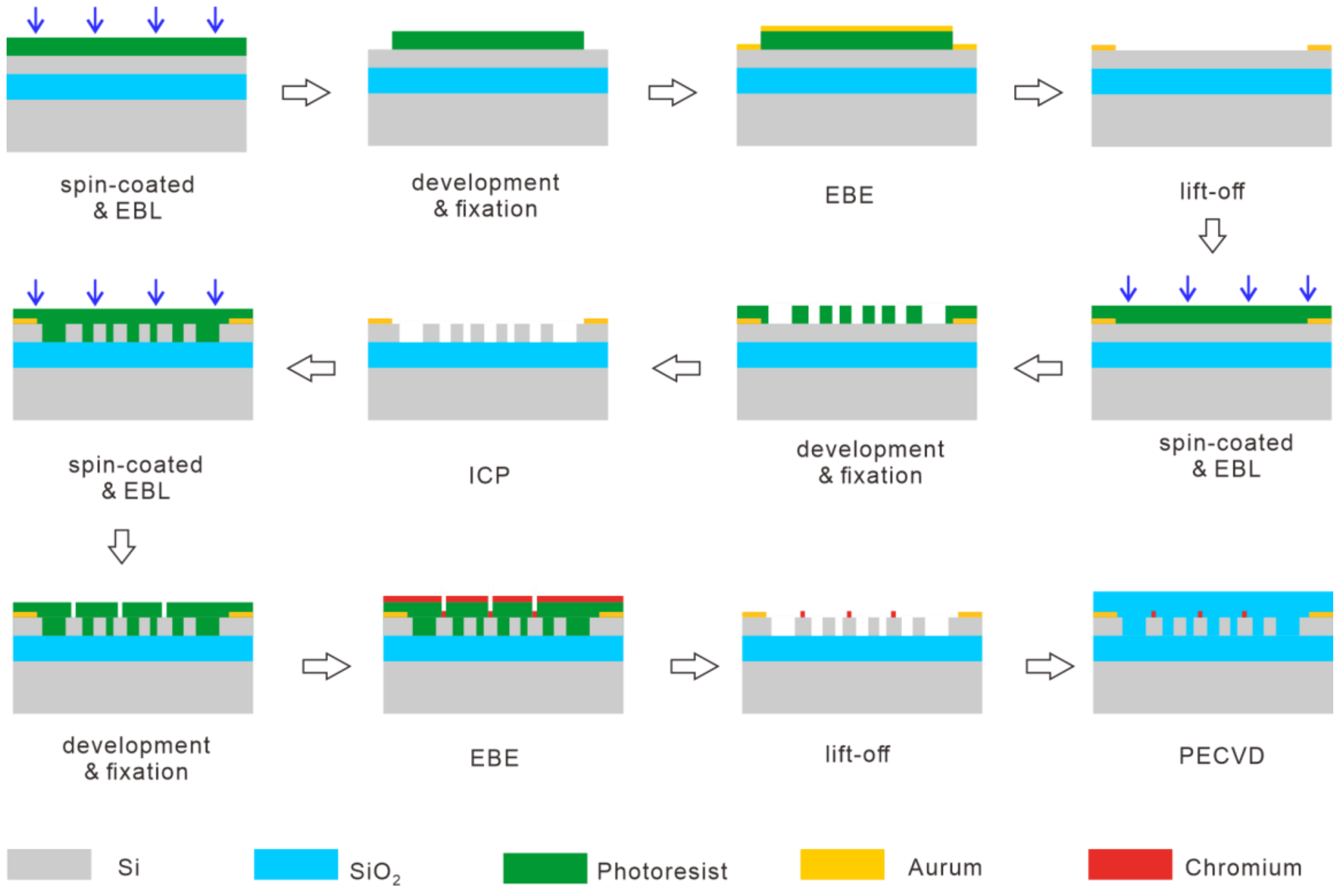

**Fig. S5.** Fabrication process of the samples.

Figure S5 shows the fabrication process of DSWs samples with a combination of three-step electron-beam lithography (EBL), inductively coupled plasma (ICP) etching, electron-beam evaporation (EBE), and plasma-enhanced chemical vapor deposition (PECVD).

Firstly, an SOI wafer was successively cleaned in ultrasound bath in acetone, isopropyl alcohol and DI water, and then was dried under nitrogen flow. The alignment marks, 20-nm-thick Aurum as an adhere layer, were fabricated by the first-step EBL, EBE and lift-off process. Photoresist was spin-coated onto the wafer surface and was patterned by EBL, which was followed by development and fixation. The Chromium and Aurum layers were successively deposited by EBE, and the final alignment marks were formed by lift-off process. Secondly, the silicon waveguides and gratings were fabricated by using a second-step EBL and ICP etching. The photoresist was patterned by use of the above-mentioned EBL process, followed by ICP etching to define waveguides and gratings. Thirdly, the Chromium layer on the first waveguides was fabricated by the third-step EBL, EBE and lift-off process. After the photoresist film was spin-coated, the pattern of Chromium is formed by EBL with careful alignment. Subsequently, a 20-nm-thick Chromium layer was deposited using EBE, followed by lift-off process to keep the required Chromium pattern. Finally, a 2-μm-thick $SiO_2$ layer is deposited by PECVD, to cover the entire sample for the optical field symmetry and structural protection.

## Supplementary Note 7: Adiabatic waveguides

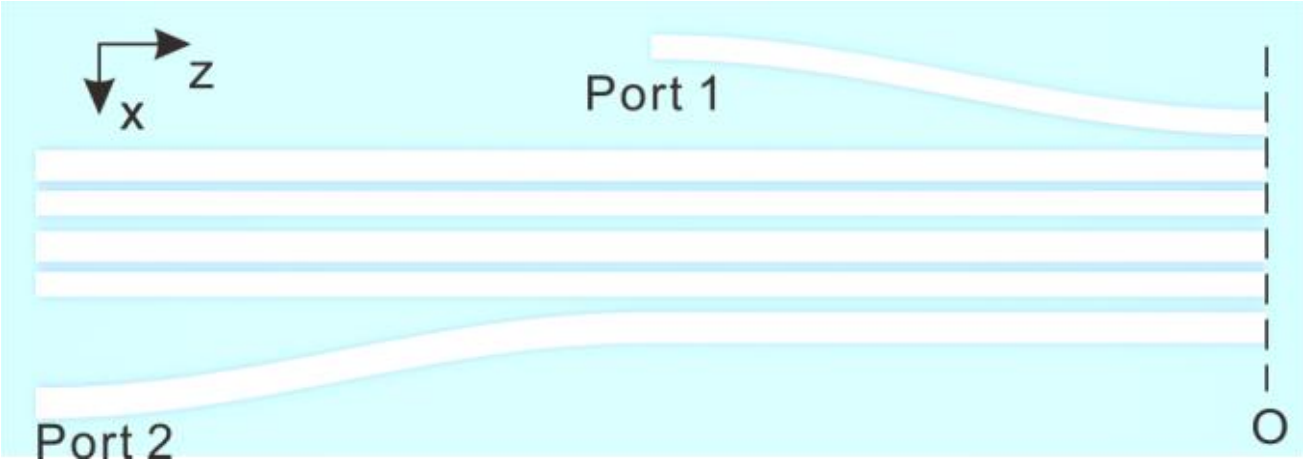

**Fig. S6.** The top view of adiabatic waveguides.

Figure S6 schematically shows the structural configuration of adiabatic waveguides at the input/output ends for the left port of the designed device. The length of adiabatic waveguides is 400 μm. The distance between port 1 and location O in the z-direction is 200 μm. The waveguides of port 1 and port 2 gradually approach waveguide arrays, which ensures that the mode evolution is sufficiently adiabatic. When the $TE_0$ mode is input from port 1 (port 2), the mode eventually evolves to become TES1 (TES2) at location O. For the device we designed, the structure of location O and location O' is the same. Therefore, the adiabatic waveguides for the left port and the right port of the designed device are mirror symmetrical.

**Supplementary Note 8: Measurement scheme**

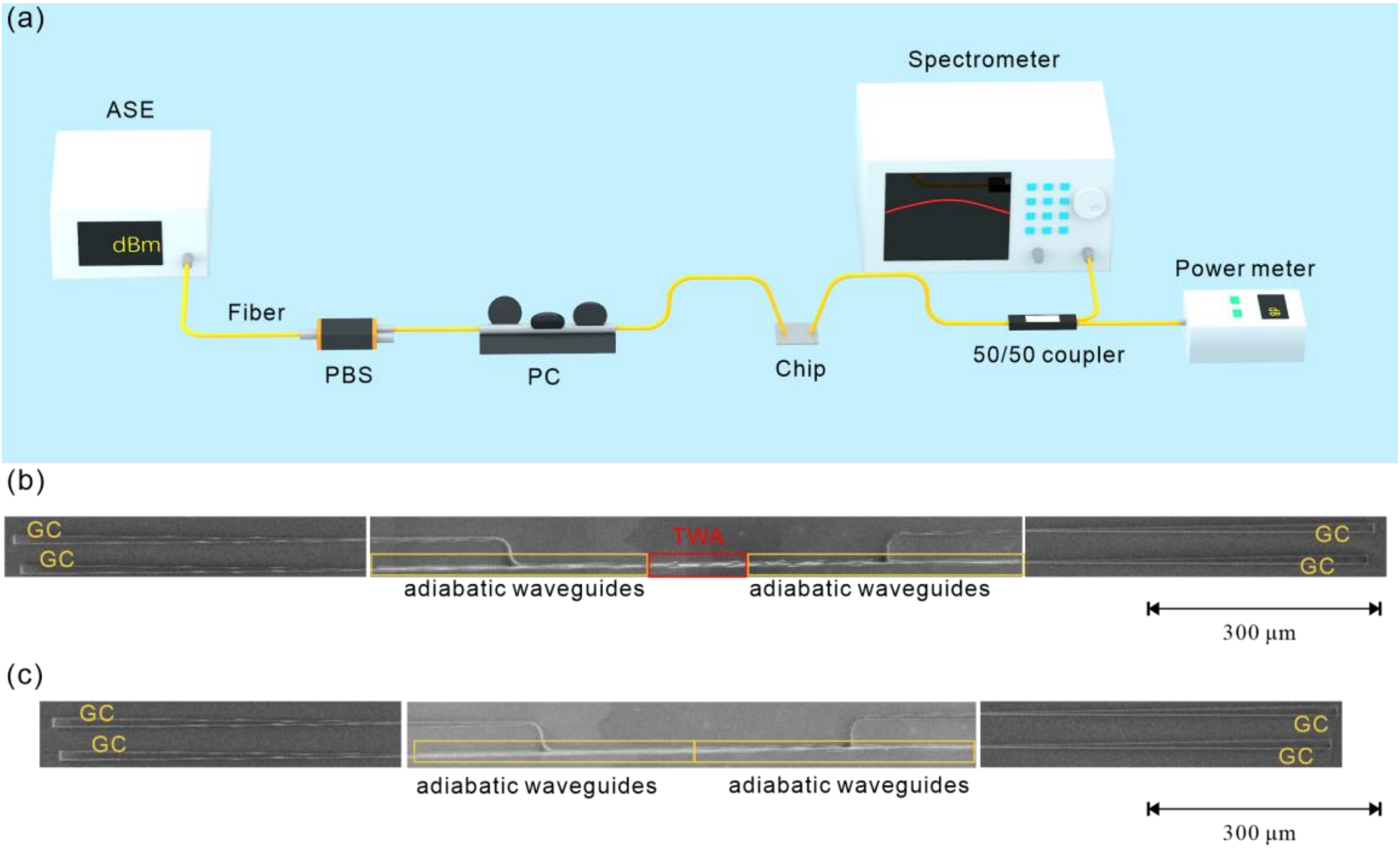

**Fig. S7.** Measurement scheme. (a) The experimental configuration. (b) The SEM image of the fabricated sample consisting of topological waveguide arrays (TWA), GCs, and adiabatic waveguides. (c) The SEM image of the control device without TWA.

Figure S7(a) presents the experimental setup for measuring the transmittance of fabricated sample. The near infrared light source is provided by an amplified spontaneous emission (ASE) broadband light source (OVLINK ASE-CL-20-B, a total power of 20 dBm, spectral range 1525 to 1600 nm). The polarization of the light source is adjusted by polarization beam splitter (PBS) and polarization controller (PC) before light is coupled into the grating coupler (GC) through the fiber. The emergent light from the SOI chip is coupled back into the fiber through the GC, and reaches 50/50 coupler, which is connected to the optical power meter and spectrometer. The optical power meter is used to adjust the angle between the fiber and the GC so as to maximize the coupling efficiency between them. The spectrometer is used to extract the transmittance for all the output modes.

Figures S7(b), S7(c) show the measured SEM images of the fabricated samples and control device, respectively. The control device without topological waveguide arrays

(TWA) is used to evaluate the loss arising from adiabatic waveguides and GCs. The transmittance at different ports can be obtained by comparing the loss differences between the fabricated sample consisting of TWA and the control device without TWA. In our measurement, we have recorded the output power in Fig. S7(b), S7(c), marked as $P_{out1}$, and $P_{out2}$, respectively, as the input power is the same, marked as $P_{in}$. The loss coefficients from the GCs, adiabatic waveguides, and TWA are assumed to be $\alpha_{GC}$, $\alpha_{AW}$, and $\alpha_{TWA}$. We can thus establish three equations associated with Figs. S7(b), S7(c),

$$P_{out1} = P_{in} \cdot e^{-(\alpha_{GC}+\alpha_{AW})} \cdot e^{-\alpha_{TWA}} \tag{S7}$$

$$P_{out2} = P_{in} \cdot e^{-(\alpha_{GC}+\alpha_{AW})} \tag{S8}$$

$$T = e^{-\alpha_{TWA}} \tag{S9}$$

Therefore, $T = P_{out1} / P_{out2}$.